\documentclass[runningheads]{svmult}

\usepackage{makeidx}   
\usepackage{graphicx}  
\usepackage{subeqnar}  
\usepackage{multicol}  
\usepackage{physprbb}  
\makeindex             

\begin{document}
\title*{Boltzmann's Approach to Statistical Mechanics}
%
%
%
%
\titlerunning{Boltzmann's Approach to Statistical Mechanics}
%
\author{Sheldon Goldstein}

%
%
%
\institute{Departments of Mathematics and Physics\\ 
Rutgers University\\ Piscataway NJ 08854, USA}

\maketitle              

\begin{abstract}
In the last quarter of the nineteenth century, Ludwig Boltzmann explained
how irreversible macroscopic laws, in particular the second law of
thermodynamics, originate in the time-reversible laws of microscopic
physics. Boltzmann's analysis, the essence of which I shall review here, is
basically correct. The most famous criticisms of Boltzmann's later work on
the subject have little merit. Most twentieth century innovations -- such
as the identification of the state of a physical system with a probability
distribution $\varrho$ on its phase space, of its thermodynamic entropy
with the Gibbs entropy of $\varrho$, and the invocation of the notions of
ergodicity and mixing for the justification of the foundations of
statistical mechanics -- are thoroughly misguided.
\end{abstract}

\section{Introduction}
I shall focus here on Boltzmann's approach to the problem of the arrow of
time: the origin of irreversible macroscopic laws, e.g., and most
importantly, the second law of thermodynamics -- the law of increasing
entropy -- in the reversible laws of microscopic physics. I shall assume,
as of course did Boltzmann, a classical framework, for the most part
ignoring quantum mechanics. 

As a matter of fact, it is widely believed that the transition from
classical mechanics to quantum mechanics requires no essential modification
of Boltzmann's ideas. This may well be so, but I believe that the
traditional formulation of quantum theory is much too vague to permit any
definitive conclusions on this matter. (For a non-traditional formulation of
quantum mechanics that avoids its conceptual incoherence, a formulation
which I quite naturally believe is well worth very serious consideration,
see the contribution of Detlef D\"urr to this volume.)

For a more detailed presentation of much of what is discussed here, the
reader should consult the excellent papers of Joel Lebowitz \cite{joel} on
this subject, as well as Jean Bricmont's sharp critique \cite{jean} of some
recent proposals. See also the contributions of Bricmont and Herbert Spohn
to this volume.

Most macroscopic phenomena are irreversible: They would look quite
different, and in fact usually quite astonishing, if run backwards in
time. Consider for example the breaking of a glass, the boiling of an egg,
birth and death. At a certain time a glass sits on a table; at another
time, a {\it later\/}  time, what had been that glass now lies on the floor
in a mess of tiny fragments. And corresponding to such irreversible
phenomena there are irreversible equations and laws, for example the
diffusion or heat equation, the Navier-Stokes equation, Boltzmann's
equation, and, perhaps most important for the issue at hand, the second law
of thermodynamics, which can be regarded as lying behind the
irreversibility of the irreversible macroscopic equations.      

This irreversible behavior and these irreversible laws must somehow be a
consequence of the (more) fundamental microscopic laws governing the
behavior of the constituents of the systems obeying the irreversible laws.
But these microscopic laws are symmetric under time reversal, and their
solutions, run backwards in time, are also solutions. The
correct detailed resolution of this apparent paradox was provided by
Boltzmann more than a century ago. And the essential idea was understood,
by Maxwell and Lord Kelvin, even earlier. Here is Lord
Kelvin~\cite{Kelvin}, writing in 1874:
\begin{quotation}
If, then, the motion of every particle of matter in the universe were
precisely reversed at any instant, the course of nature would be simply
reversed for ever after. The bursting bubble of foam at the foot of a
waterfall would reunite and descend into the water \dots\ Boulders would
recover from the mud the materials required to rebuild them into their
previous jagged forms, and would become reunited to the mountain peak from
which they had formerly broken away. And if the materialistic  hypothesis
of life were true, living creatures would grow backwards, with conscious
knowledge of the future, but no memory of the past, and would become again
unborn. But the real phenomena of life infinitely transcend human science
\dots\ Far otherwise, however, is it in respect to the reversal of the
motions of matter uninfluenced by life, a very elementary consideration of
which leads to a full explanation of the theory of dissipation of energy. 
\end{quotation}

The adequacy of Boltzmann's resolution was controversial at the time and
remains so even today. Nonetheless, the detailed solution found by Boltzmann,
based in part on his understanding of the microscopic meaning of entropy,
was well appreciated and admired by many of his contemporaries. For
example, towards the end of his life Schr\"odinger \cite{sc1} reported that
\lq \lq no perception in physics has ever seemed more important to me than that
of Boltzmann -- despite Planck and Einstein.'' In a more detailed
assessment, Schr\"odinger \cite{sc2} declared that
\begin{quotation}
\noindent The spontaneous transition from order to disorder is the
quintessence of Boltzmann's theory \dots\ This theory really grants an
understanding and does not \dots\ reason away the dissymetry of things by
means of an a priori sense of direction of time \dots\ No one who has once
understood Boltzmann's theory will ever again have recourse to such
expedients. It would be a scientific regression beside which a repudiation
of Copernicus in favor of Ptolemy would seem trifling.
\end{quotation}

However, as I've already indicated, the issue has never been entirely
settled to everyone's complete satisfaction. In fact Schr\"odinger
concluded the statement that I've just quoted with the observation that
\begin{quotation}
\noindent Nevertheless, objections to the theory have been raised again and
again in the course of past decades and not (only) by fools but (also) by fine
thinkers. If we \dots\ eliminate the subtle misunderstandings \dots\ we
\dots\ find \dots\ a significant residue \dots\ which needs exploring \dots\ 
\end{quotation}
Notice Schr\"odinger's parenthetical comments, suggesting that he did not
have a very high opinion of most of the objections that continued to be
raised against Boltzmann's ideas. Note also his separation of a
residual core of confusion from what was regarded -- or should have been -- as
settled. 

Roughly corresponding to this separation, the problem of irreversibility can
be regarded as having two parts: an easy part and a hard part. For many
physicists who find Boltzmann's solution inadequate, it is only the
\lq \lq hard part'' that causes them trouble.  However, for a great many
others even the \lq \lq easy part'' seems hard, and generates an enormous
amount of confusion. 

Moreover, even with the solution to both parts of the problem, there are
still questions, of a philosophical character, that remain. These
questions, touched upon in Section 4, could be regarded as defining the
really hard part of the problem. But it would be better to see them, not so
much as part of the problem of irreversibility, but rather as general
questions about the meaning and nature of scientific explanation.

\section{The Easy Part}
Boltzmann's great achievement was to arrive at an understanding of the
meaning of entropy and why it should tend to increase, almost never
decreasing. This understanding is such as also to explain why systems
evolve to equilibrium states, of maximal entropy. 

Consider, for example the sequence of snapshots of a gas in a box
illustrated in Fig.~1.  
\begin{figure}[h]
\begin{center}
\includegraphics[width=.9\textwidth]{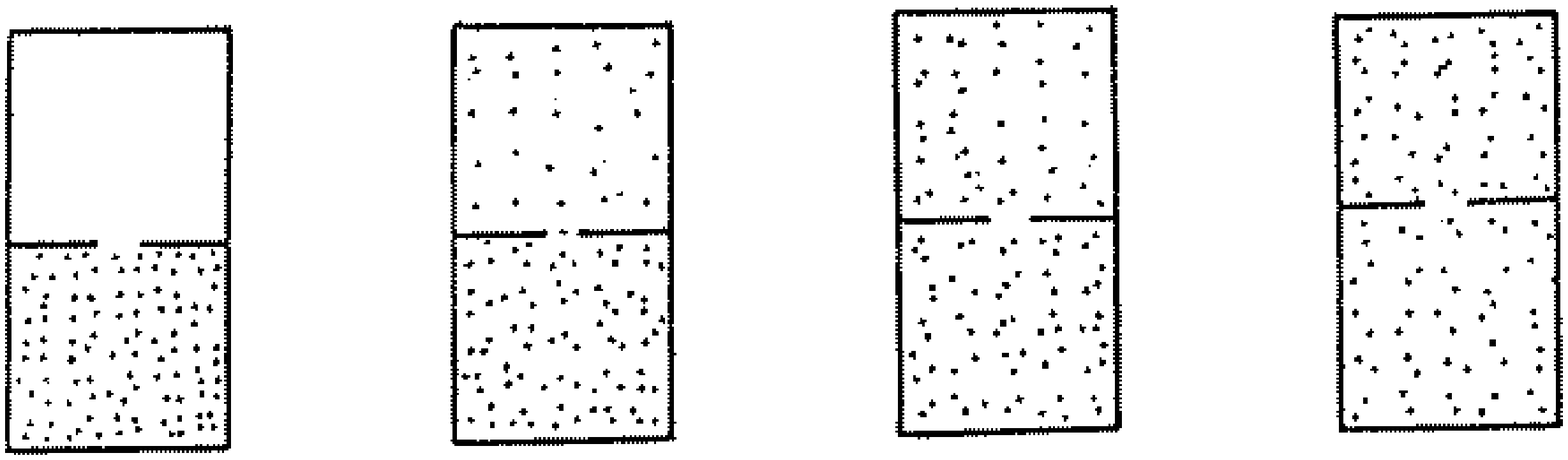}
\end{center}
\caption[]{Snapshots of a gas in a box}
\label{eps1}
\end{figure}
We see here the transition from a low entropy nonequilibrium state on the
left, with the gas entirely on the bottom of the box, through states of
higher entropy, with the gas expanding into the full box, until at the
right we have a high entropy equilibrium state with the gas uniformly
distributed throughout the box.

The question that Boltzmann addressed is quite simple and straightforward:
Why does the gas prefer to be in equilibrium -- that is, to look like the
snapshot on the right? The answer that he found is, perhaps, even simpler
but also, somehow, rather subtle.

A complete description of the state of the gas is provided by its {\it
phase point\/} $X$, a point in the phase space of possible microscopic
states of the gas.  Many different phase points correspond to each of the
snapshots in Fig.~1. There is nothing particularly special about any
specific {\it equilibrium phase point\/} -- a phase point corresponding to
the snapshot on the right, a system in equilibrium. The dynamics prefers
a given equilibrium phase point neither more nor less than it prefers any
other given phase point, even a specific far-from-equilibrium phase
point, corresponding say to the leftmost snapshot. 

There are, however, for a system at a given energy $E$, far more
equilibrium phase points than nonequilibrium phase points, overwhelming
more, in fact, than the totality of nonequilibrium phase points at that
energy -- corresponding to all possible ways the system can fail to be in
equilibrium, and described, for example, by the various density
distributions perceptibly different from the uniform one. The only relevant
sense in which the equilibrium phase points could be regarded as special is
that there are vastly more of them.

I shall now sketch Boltzmann's account of why this should be so, indicating
the central role of entropy and its meaning in providing a quantitative
foundation for what I've just described. A crucial ingredient in this
account is the vast separation of scales between the microscopic and the
macroscopic levels, without which irreversible phenomena could not possibly
emerge from reversible microscopic laws. 

\subsection*{Boltzmann's Entropy}
 
Consider the microscopic state $X=\left(\vec{q_1},
\dots,\vec{q_N},\vec{p_1}, \dots,\vec{p_N}\right) $ of a classical system
consisting of a large number $N$ of identical particles forming a gas in a box
$\Lambda$, with positions $\vec{q_i}\in\Lambda$ and momenta
$\vec{p_i}$. The evolution of the system is determined, via Hamilton's
equations of motion $\D\vec{q_i}/\D t=\partial H/\partial\vec{p_i},\
\D\vec{p_i}/\D t=-\partial H/\partial\vec{q_i}$, by a Hamiltonian function
$H(\vec{q_1}, \dots,\vec{q_N},\vec{p_1}, \dots,\vec{p_N})$, the energy
function for the system, normally assumed to be the sum of a kinetic energy
term and a potential energy term but whose form shall not much concern us
here. Since the energy is a constant of this motion, we may take as the
relevant set of possible states, not the full phase space $\Omega(\Lambda)$
of the system under consideration but the energy surface
$\Omega_E=\left\{X\in\Omega(\Lambda)\,|\,H(X)=E\right\}$ corresponding to
the value $E$ of the energy of the system.

Each snapshot in Fig.~1 corresponds to a subset $\Gamma$ of $\Omega_E$,
namely the set of all phase points that \lq\lq look like'' the
snapshot. Every phase point $X$ belongs to a macrostate $\Gamma(X)$
consisting of phase points that are macroscopically similar to $X$. More
precisely, partition the 1-particle phase space (the $\vec{q},
\vec{p}$ -- space) into macroscopically small but microscopically large cells
$\Delta_\alpha$ and specify (approximately, by providing intervals) the
number $n_\alpha$ of particles in each cell. Each such specification determines
a {\it macrostate\/}, the set of all phase points whose particles are
distributed in the manner specified. Different specifications yield
different macrostates and the set of all such macrostates defines a
partition  of our phase space $\Omega_E$, into macrostates. $\Gamma(X)$ is
then the macrostate -- the element of this partition -- to which $X$
belongs. All points in $\Gamma(X)$ have, in particular, similar spatial
particle densities (and  similar bulk-velocity profiles).

Different macrostates typically have vastly different sizes, and this
difference is conveniently quantified by  {\it Boltzmann's entropy\/}:

\begin{equation}\label{entropy}
S(X)=k\log\left|\Gamma(X)\right|\;,
\end {equation} 
where $k$ is Boltzmann's constant, expressing the relationship between
macroscopic and microscopic units, and $\left|\ \ \right|$ denotes volume,
given by the time-invariant projection of the Liouville (Lebesgue) measure
onto $\Omega_E$. (The definition (\ref{entropy}) is correct up to an
additive constant, depending on $N$, that is important but shall not
concern us here.) Following Boltzmann, let us identify Boltzmann's entropy
with the thermodynamic entropy, to which the Second Law refers. Then that
we should expect entropy to increase -- but for exceedingly rare exceptions
never decreasing -- more or less directly follows from the very concept
(\ref{entropy}) of entropy once the vast separation of scales between the
microscopic and the macroscopic levels is taken into account.

This is because the entropy $S$ is an extensive variable, proportional to
the number $N$ of particles in the system. In fact, the \lq \lq log'' in
(\ref{entropy}) suggests that the various possible values of
$\left|\Gamma(X)\right|$ need to be measured on a logarithmic scale, these
values differing so greatly that without taking the logarithm it would be
difficult to place them on the same scale or graph. With the logarithm, an
extensive quantity is obtained, and after dividing by the spatial volume of
the system we obtain an entropy per unit volume whose differences, for
different macrostates, are of order unity.

What all of this means is that typical values of $\left|\Gamma(X)\right|$,
and, more to the point, typical ratios of such values, for example between
the value of $\left|\Gamma(X)\right|$ for the equilibrium state depicted in
the snapshot on the right of Fig.~1 and the value for any nonequilibrium
state, say the one depicted on the left, are of \lq \lq order'' $10^N$, or,
more explicitly, of order $10^{10^{20}}$ for a system with $10^{20}$
particles. (That this is the correct order is easily seen by noting that
when the condition that all particles be on the left side of the box is
dropped, the volume of the set of all possible phase points consistent with
the constraints increases by a factor of $2^N$ -- a factor of 2 for each of
the particles. Note also that in terms of the sense of \lq \lq order''
appropriate for comparing quantities as vast as those under consideration
here, \lq \lq of order $10^{10^{20}}$'' should not be regarded as different
from \lq \lq of order $2^{10^{20}}$.'')

Thus $\Omega_E$ consists almost entirely of phase points in the equilibrium
macrostate $\Gamma_{\mathrm{eq}}$, with ridiculously few exceptions whose
totality has volume of order $10^{-10^{20}}$ relative to that of
$\Omega_E$. For a nonequilibrium phase point $X$ of energy $E$, the
Hamiltonian dynamics governing the motion $X_t$ arising from $X$ would have
to be ridiculously special to avoid reasonably quickly carrying $X_t$ into
$\Gamma_{\mathrm{eq}}$ and keeping it there for an extremely long time --
unless, of course, $X$ itself were ridiculously special.

\subsection*{The Objections of Zermelo and Loschmidt}

Zermelo and Loschmidt claimed that Boltzmann's analysis, showing how
irreversible macroscopic behavior emerges from reversible microscopic laws,
was -- and had to be -- inadequate. Loschmidt noted that since the
classical equations of motion are time-reversible, we may obtain solutions
$X_t$ to these equations that violate the macroscopic laws by
time-reversing solutions that obey them, thereby obtaining, for example, a
solution which runs through the snapshots in Fig.~1 in time-reversed order,
right to left. And Zermelo pointed out that there is another reason why
such anti-thermodynamic solutions must exist: Poincar\'e recurrence, which
guarantees that most solutions (the exceptions have, in fact, at most
measure 0) that initially belong to the macrostate depicted in the left
snapshot will eventually return to that macrostate. It thus follows that
anti-thermodynamic behavior is just as consistent with the microscopic laws as
thermodynamic behavior (which is of course obvious in view of the
reversibility of those laws), so that neither, in fact, could be a
consequence of those laws.

However, no genuine conflict with the analysis of Boltzmann follows from
these objections. Boltzmann did not (finally) claim that all phase points
should behave properly, but only that by far most of them -- in any given
macrostate and in the sense of volume relative to that of the macrostate --
should. Some may behave improperly, and the arguments of Zermelo and
Lo\-schmidt merely show that such bad phase points in fact exist. For
example, from Loschmidt we know that we may obtain such points by reversing
the velocities of the equilibrium phase points to which nonequilibrium
phase points (reasonably quickly) evolve.

Here is part of Boltzmann's response to Zermelo \cite{b}:
\begin{quotation}
\noindent I have \dots\ emphasized that the second law of thermodynamics is
from the molecular viewpoint merely a statistical law. Zermelo's paper
shows that my writings have been misunderstood; \dots\ Poincar\'e's
theorem, which Zermelo explains at the beginning of his paper, is clearly
correct, but his application of it to the theory of heat is not. \dots\
Thus, when Zermelo concludes, from the theoretical fact that the initial
states in a gas must recur -- without having calculated how long a time
this will take -- that the hypotheses of gas theory must be rejected or
else fundamentally changed, he is just like a dice player who has
calculated that the probability of a sequence of 1000 one's is not zero,
and then concludes that his dice must be loaded since he has not yet
observed such a sequence!
\end{quotation}

\subsection*{The Relevance of Ergodicity and Mixing}

The basic notions of ergodic theory, in particular ergodicity and mixing,
are widely believed to play a crucial role in the foundations of
statistical mechanics. Ergodicity, roughly speaking the absence of
constants of the motion other than functions of the energy $H$, implies the
equality of time-averages and phase-space averages, i.e., of the long-time
average of a quantity as it changes under the dynamics and the uniform
average of that quantity, over the relevant energy surface $\Omega_E$, as
the phase point at which it is evaluated varies over that surface. This
is supposed to justify the use, to define the equilibrium values of
thermodynamic quantities, of such phase-space averages with respect to the
{\it microcanonical ensemble\/}, the uniform distribution over the energy
surface, the idea being that the observed values of these quantities are
time-averages, since measurement takes time.

This use of ergodicity is thoroughly misguided. Boltzmann's key insight was
that, given the energy of a system, the overwhelming majority of its phase
points on the corresponding energy surface are equilibrium phase points,
all of which look macroscopically more or less the same. This means that
the value of any thermodynamic quantity is, to all intents and purposes,
constant on the energy surface, and averaging over the energy surface will
thus reproduce that constant value, regardless of whether or not the system
is ergodic.

For example, one characteristic shared by the equilibrium phase points, and
thus by the great majority of phase points on the energy surface, is a
common Maxwellian empirical distribution for the velocities.  Here again is
Boltzmann \cite{b}, expressing his frustration with Zermelo for failing, it
seems, to appreciate that fact:
\begin{quotation}
\noindent [The Maxwell distribution] is characterized by the fact
that by far the largest number of possible velocity distributions have the
characteristic properties of the Maxwell distribution, and compared to
these there are only a relatively small number of possible distributions
that deviate significantly from Maxwell's. Whereas Zermelo says that the
number of states that finally lead to the Maxwellian state is small
compared to all possible states, I assert on the contrary that by far the
largest number of possible states are ``Maxwellian'' and that the number that
deviate from the Maxwellian state is vanishingly small.
\end{quotation} 

There is another problem with this use of ergodicity, a mismatch of time
scales. The time scale appropriate for the ergodicity of a gas in a box is,
roughly speaking, the time necessary for a trajectory for the motion in
phase space to wind all over the phase space, and this is at least as long
as a Poincar\'e recurrence time, the time necessary for the gas, say after
leaving the state depicted in the left snapshot in Fig.~1, to return to
that state, a time as large as the corresponding microstate is small and
hence of order $10^{10^{20}}$ in your favorite unit of time, a time far far
larger than that believed to be the age of the universe, since the big
bang. Thus ergodicity couldn't possibly be very relevant to an account of
phenomena, such as those with which thermodynamics is concerned, taking
place on reasonable time scales.

Mixing is supposed to explain why systems evolve to a state of
equilibrium. The idea here is that since such a state is in a sense
characterized by a special probability distribution, namely the microcanonical
ensemble, evolution to equilibrium amounts to the convergence of a generic
(nonequilibrium) distribution to the special one, under the dynamics on
probabilities arising from the phase-space motion. But this, made suitably
precise, amounts to the assertion that the system is (a) {\it mixing\/}
(system), this terminology referring to the fact that the notion is usually
defined by the following (equivalent) property of the dynamics: The points of
any region $R$ -- of nonvanishing volume, no matter how small -- of the
energy surface will evolve under the dynamics, after a suitably long time $t$,
to points that fill out a region $R_t$ that is distorted and convoluted in
such a way as to be spread more or less uniformly throughout the energy
surface, in the sense that for any reasonable function $f$, the uniform
averages of $f$ over $\Omega_E$ and over $R_t$ are more or less the same.

Since the energy surface $\Omega_E$ consists almost entirely of a single
macrostate, the equilibrium macrostate $\Gamma_{\mathrm{eq}}$, the mixing
property of the evolution on the energy surface pretty much amounts to the
condition that (even small) subregions of $\Gamma_{\mathrm{eq}}$ become
uniformly spread over $\Gamma_{\mathrm{eq}}$ after a sufficiently long
time. But this could not possibly be relevant to the problem of approach to
equilibrium, since it concerns only phase points that describe a system
that is already in equilibrium.  Approach to equilibrium concerns the
passage of a nonequilibrium phase point, lying outside of
$\Gamma_{\mathrm{eq}}$, into $\Gamma_{\mathrm{eq}}$, and this tends to
happen, in fact typically rather quickly, merely because
$\Gamma_{\mathrm{eq}}$ is so extremely large. (Note, however, that although
mixing is a much stronger property of the dynamics than ergodicity, the
mixing time scale could be much smaller than that for ergodicity, and could
in fact be of the same order as the time required for a system to reach
equilibrium.) 

This abuse of mixing is so obviously wrong-headed that one can't help
wondering about the sources of the confusion. Here are two possibilities:

One of the most widely mentioned metaphors for mixing, invoked by Gibbs
\cite{gibbs} when he introduced the notion, is the spreading of a tiny ink
drop when it is placed in a glass of water and the water is stirred. The
spreading of the ink all over the water is also an example of approach to
equilibrium. However, it is important to bear in mind that insofar as it
illustrates mixing, this example should be regarded as having as its phase
space the points of the liquid, a three-dimensional set, whereas the
relevant phase space for the approach to equilibrium in the example is the
set of configurations of the liquid, describing how the ink is distributed
in the water, a space of enormous dimension. The sort of mixing that can be
relevant to approach to equilibrium takes place in physical space, not in
phase space.

A related point: there is a quantity of a probablistic character whose
approach to equilibrium for a low density gas does reflect, famously and
correctly, that of the system. I have in mind here, of course, the
one-particle Boltzmann function $f(\vec{q}, \vec{p})$, whose approach to
equilibrium is governed by Boltzmann's equation and which describes the
empirical distribution, or coarse-grained density, of the particles of the
gas in the one-particle phase space. It is worth stressing that
$f(\vec{q},\vec{p})$, since it is an empirical distribution, is determined
by the state of the system, given by a point $X$ in its full phase
space. $f(\vec{q},\vec{p})$ can in fact be regarded as describing a
macrostate of the system by using it to specify, up to a constant of
proportionality, the numbers $n_\alpha$ defining the macrostate; see the
paragraph preceding the one containing equation \ref{entropy}.

This legitimate association of approach to equilibrium for the system with
the approach of $f$ to an equilibrium distribution, Maxwellian in velocity
and spatially homogeneous, has unfortunately suggested to many that
approach to equilibrium for a more general gas should be identified with
the convergence of the full $n$-particle distribution function to one that
is constant on the relevant energy surface, describing the microcanonical
ensemble -- in a word, mixing. But the $n$-particle distribution function is
not an empirical distribution and, unlike $f(\vec{q},\vec{p})$, is not
merely a characteristic of the actual state $X$ of the gas.

Another source of confusion lies in the widespread tendency to identify
states of a physical system with probability measures on its phase space, a
tendency partly owing to the very success achieved by statistical mechanics
through the use of statistical methods in general, and the standard
ensembles in particular; and partly owing to the baleful influence of
quantum mechanics, one of the main lessons of which is all too widely
believed to be that a detailed description of a quantum mechanical system
is fundamentally impossible, so that the state of such a system must be
identified with an object of a statistical character, be it the wave
function of the system or a positive linear functional on its algebra of
observables.  

\subsection*{Boltzmann's Entropy Versus the Gibbs Entropy}
The identification of the state of a system with a probability measure,
given, say, by a density $\varrho$ on its phase space, has led to the
widespread identification of the thermodynamic entropy of a system with its
{\it Gibbs entropy\/}:
\begin{equation}\label{gibbs}
S_{\mathrm{G}}(\varrho)=-k\int\varrho(X)\log\varrho(X)\,\D X\;.
\end {equation} 
One of the most important features of the Gibbs entropy is that it is a
constant of the motion: Writing $\varrho_t$ for the evolution on densities
induced by the motion on phase space, we have that
$S_{\mathrm{G}}(\varrho_t)$ is independent of $t$; in particular it does
not increase to its equilibrium value.

It is frequently asked how this can be compatible with the Second Law.  The
answer is very simple. The Second Law is concerned with the thermodynamic
entropy, and this is given by Boltzmann's entropy (\ref{entropy}), not by
the Gibbs entropy (\ref{gibbs}). In fact, the Gibbs entropy is not even an
entity of the right sort: It is a function of a probability distribution,
i.e., of an ensemble of systems, and not a function on phase space, a
function of the actual state $X$ of an individual system, the behavior of
which the Second Law -- and macro-physics in general -- is supposed to describe.

The widespread tendency to identify the thermodynamic entropy with the Gibbs
entropy rather than with Boltzmann's entropy, while clearly misguided, is
easy to understand. Boltzmann's entropy is a bit vague, since it depends
upon somewhat arbitrary choices that go into the specification of the
macrostates of a system, and, other things being equal, vagueness is
bad. This vagueness, however, is of little practical consequence, and
indeed upon reflection is quite appropriate for the problem of defining
entropy, a concept that somehow relates the microscopic level of
description and the (not sharply definable) macroscopic level. But the
vagueness is there nonetheless.

Unlike Boltzmann's entropy, the Gibbs entropy is sharply defined, without
arbitrariness. It is also a very natural functional of a probability
distribution, having a great many useful and appealing features into which I
shall not enter here. It is the basis of extremely important
developments in information theory and in the ergodic theory of dynamical
systems, and has proven to be of great value in the derivation of
hydrodynamical laws from suitable microscopic first principles. It is a very
fine concept and can be used to define a great many things, but
thermodynamic entropy is not one of them.

Certainly contributing to the tendency to identify the thermodynamic
entropy with the Gibbs entropy is the fact that for systems in equilibrium
the Gibbs entropy agrees with Boltzmann's entropy. More precisely, and
more generally, $S(X)=S_{\mathrm{G}}(\varrho)$ whenever $\varrho$ is the
uniform distribution over the macrostate to which $X$
belongs. Moreover -- and this is probably the origin of the
confusion -- Boltzmann showed that for a low density gas
\begin{equation}\label{bg}
S(X)\approx -kN\int f(\vec{q}, \vec{p})\ln f(\vec{q}, \vec{p})\,\D \vec{q}
\D \vec{p}
\end{equation}
whenever $X$ belongs to the macrostate defined by the Boltzmann function
$f(\vec{q}, \vec{p})$. 

This formula involves the Gibbs entropy for the one-particle distribution
$f$ and is appropriate, as mentioned, for a low density gas, in which
correlations between particles are insignificant. This has suggested to
many that to obtain an expression for entropy valid beyond the low density
regime one should apply a similar formula to the full probability
distribution, capturing all the correlations then present. The result, of
course, is the Gibbs entropy (\ref{gibbs}). The mistake in doing so lies in
failing to appreciate that it is the left hand side of (\ref{bg}), defined
by (\ref{entropy}), that is fundamental, not the right hand side, which is
merely the result of a computation valid in a special situation.

It is widely believed that thermodynamic entropy is a reflection of our
ignorance of the precise microscopic state of a macroscopic system, and
that if we somehow knew the exact phase point for the system, its
entropy would be zero or meaningless. But entropy is a quantity playing a
precise role in a formalism governing an aspect of the behavior of
macroscopic systems. This behavior is completely determined by the evolution
of the detailed microscopic state of these systems, regardless of what any
person or any other being happens to know about that state. The widespread
acceptance of the preposterous notion that how macroscopic systems behave
could be affected merely by what we know about them is simply another
instance of the distressing effect that quantum mechanics has had upon the
ability of physicists to think clearly about fundamental issues in physics.

\section{The Hard Part}
In the previous section we addressed the easy part of the problem of
irreversibility: Suppose a system, e.g., a gas in a box, is in a state of
low entropy at some time. Why should its entropy tend to be larger at a
later time? The reason is basically that states of large entropy correspond
to regions in phase space of enormously greater volume than those of lower
entropy. We now turn to the hard part of the problem: Why should there be
an arrow of time in our universe, governed as it is, at the fundamental
level, by reversible microscopic laws?

The problem here can be appreciated by focusing on the question: What is
the origin of the low entropy initial states? If they are so ``unlikely,''
why should systems find themselves in such states? In many cases, the
answer is that we or an experimenter created them, from states of lower
entropy still. If we continue to ask such questions, we come to the
conclusion that the cause of low entropy states on earth, the source in
effect of negative entropy, is our sun, whose high energy photons are
absorbed by the earth, which converts them to a great many low energy
photons (having together much larger entropy), permitting
entropy-decreasing processes to occur on our planet without violation of
overall entropy non-decrease. And if we push further we eventually arrive
at a cosmological low entropy state, in the distant past, for the universe
as a whole.

And what about the origin of this state? Figure~2, taken from Roger
Penrose's {\it The Emperor's New Mind\/} \cite[page 343]{enm}, illustrates
the difficulty.
\begin{figure}[h]
\begin{center}
\includegraphics[width=.8\textwidth]{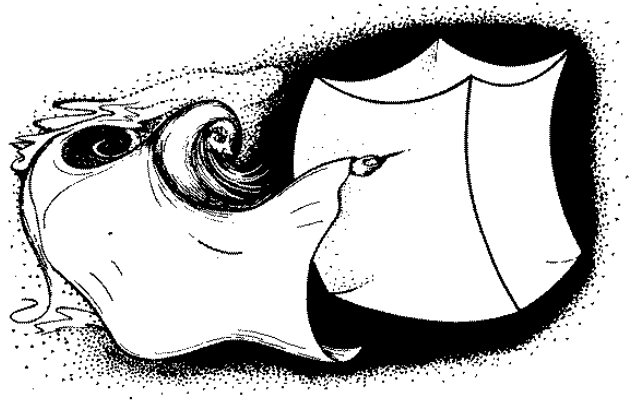}
\end{center}
\caption[ ]{In order to produce a universe resembling the one in which we
live, the Creator would have to aim for an absurdly tiny volume of the
phase space of possible universes -- about  $1/10^{10^{123}}$ of the entire
volume, for the situation under consideration. (The pin, and the spot aimed
for, are not drawn to scale!)} 
\label{eps2}
\end{figure}
Penrose estimates the volume of the region of phase space
corresponding to the possible initial states of the universe to be one part in
$10^{10^{123}}$ of the entire relevant phase space. He bases this upon the
Bekenstein-Hawking entropy of a $10^{80}$ baryon mass black hole, which,
in \lq natural units,' is ${10^{123}}$, corresponding to a \lq \lq Big
Crunch state' of volume of order $10^{10^{123}}$, a reasonable estimate,
Penrose argues, for the volume of the entire relevant phase space for a
closed universe. (It does not much matter what we take as the entropy of
the initial state, be it ${10^{20}}$ or ${10^{80}}$ or ${10^{100}}$ -- the
relevant ratio of volumes will always be of order $10^{10^{123}}$. As to
whether $10^{10^{123}}$ is indeed a good estimate for the volume of the
relevant phase space, it has been suggested that $\infty$ might be more on
target; see the contribution of Michael Kiessling to this volume. )

As to why the universe should have begun in such an exceedingly improbable
macrostate, an answer that has often been suggested is that such a state
arose from a fluctuation out of equilibrium. In fact, if the universal
dynamics were ergodic, such a fluctuation must eventually occur,
repeatedly, for all phase points with the possible exception of a set of
measure 0.

Nonetheless, this answer is quite unsatisfactory; indeed, according to
Feynman \cite{Feynman} it is \lq \lq ridiculous.'' The problem is that if the
explanation of entropy increase and the arrow of time in our universe is
that they have emerged from a low entropy state that arose from a
fluctuation, then that fluctuation should have been no larger than
necessary -- that is, to a state like the present state of the universe, and
not to a state of much lower entropy as seems to have existed in the past. Here
is Feynman, referring to astronomy, to history books and history, and to
paleontology:    
\begin{quotation}
\noindent Since we always make the prediction that in a place where we have not
looked we shall see stars in a similar condition, or find the same statement
about Napoleon, or that we shall see bones like the bones that we have seen
before, the success of all those sciences indicates that the world did not
come from a fluctuation \dots Therefore I think it is necessary to add to
the physical laws the hypothesis that in the past the universe was more
ordered \dots\ than it is today -- I think this is the additional statement
that is needed to make sense, and to make an understanding of the
irreversibility.
\end{quotation}
The view expressed here by Feynman may seem to be the view at which Boltzmann
himself ultimately arrived:
\begin{quotation}
\noindent The second law of thermodynamics can be proved from the
mechanical theory if one assumes that the present state of the universe, or
at least that part which surrounds us, started to evolve from an improbable
state and is still in a relatively improbable state. This is a reasonable
assumption to make, since it enables us to explain the facts of experience,
and one should not expect to be able to deduce it from anything more
fundamental.
\end{quotation}
However,  this statement, with which Boltzmann began his second response
to Zermelo  \cite{b1}, probably should not be read as a repudiation of the
fluctuation hypothesis ridiculed by Feynman, since towards the end of the
very same article, in its \S 4, Boltzmann advocated this hypothesis.

Be that as it may, what we need to complete Boltzmann's account of
irreversibility is a reasonable hypothesis on the initial state of the
universe, in effect an additional physical law. This hypothesis must imply
that this state had very low entropy, but, unlike Feynman's suggestion
above, it need not explicitly stipulate that this be so. What is required
is that the initial state not be too contrived, that it be somehow
reasonable, indeed that it be natural.

Moreover, it seems that gravity, even classical gravity, affords just such
a possibility. This is because the attractive nature of the gravitational
interaction is such that gravitating matter tends to clump, clumped states
having larger entropy. (For more on this see the contribution of Michael
Kiessling to this volume.)  This important difference from the behavior of
ordinary matter, for which gravity can be more or less ignored, is well 
illustrated  by Penrose in \cite[page 338]{enm}, see Fig.~3.
\begin{figure}[h]
\begin{center}
\includegraphics[width=.8\textwidth]{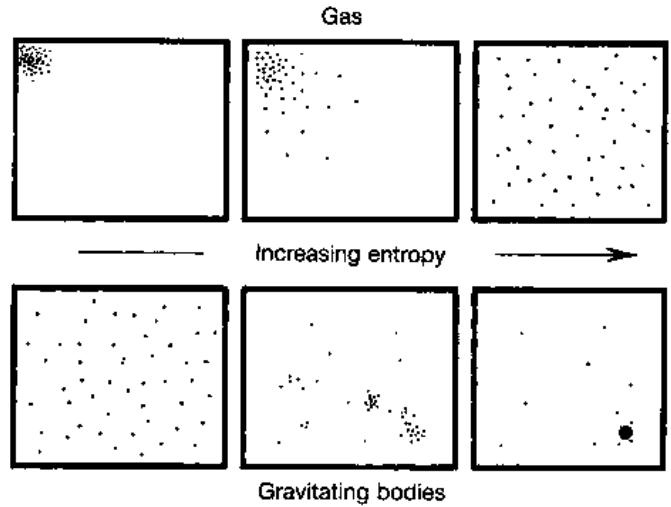}
\end{center}
\caption[]{For an ordinary gas, increasing entropy tends to make the
distribution more uniform. For a system of gravitating bodies the reverse
is true. High entropy is achieved by gravitational clumping -- and the
highest of all, by collapse to a black hole.} 
\label{eps3}
\end{figure}
What is important for our purposes here is that what is arguably the most
random, typical, natural, and least contrived initial state for a system of
gravitating particles, one in which  they are uniformly distributed over
space (with, say, all of them at rest) also happens to be a state of very
low entropy, exactly what is needed to complete Boltzmann's account of
irreversibility.

\section{Typicality and Explanation}

According to the Boltzmannian scenario propounded here, the overwhelming
majority, as measured by relative phase-space volume, of phase points in a
(very small) initial macro\-state of the universe evolve in such a way as
to behave -- for reasonable times, that are not too large on the time scale
defined by the present age of the universe, since the big bang --
thermodynamically, with suitably closed subsystems having increasing
entropy, exhibiting irreversible behavior, and so on. In other words, {\it
typical \/}phase space points yield the behavior that it was our (or
Boltzmann's) purpose to explain. Thus we should expect such behavior to be
prevail in our universe.

This raises the question, What is the force of such an explanation, based,
as it is, merely on what would ``typically'' happen, not on what must
inevitably happen? Now, as a scientist I see no problem here. What more
could reasonably be expected by way of explanation? Boltzmann \cite{b1}
makes a similar point when he complains that
\begin{quotation}
\noindent The applicability of probability theory to a particular case
cannot of course be proved rigorously. \dots\ Despite this, every insurance
company relies on probability theory. \dots\ The assumption that these rare
cases are not observed in nature is not strictly provable (nor is the
entire mechanical picture itself) but in view of what has been said it is so
natural and obvious, and so much in agreement with all experience with
probabilities, \dots that any doubt on this point certainly cannot put in
question the validity of the theory when it is otherwise so useful.
  
It is completely incomprehensible to me how anyone can see a refutation of
the applicability of probability theory in the fact that some other
argument shows that exceptions must occur now and then over a period of
eons of time; for probability theory itself teaches just the same thing.
\end{quotation}

However, as philosophers we might in fact be inclined to demand more. We
might ask whether we've really explained a phenomenon (such as
irreversibility, an arrow of time in our world, or whatever) if we have
(merely!) shown that it is typical, but that exceptions, while in a sense
extraordinarily rare, still exist in abundance -- at least as logical and
mathematical possibilities. But in doing so it seems to me that we will
have to face very hard questions about just what is meant by scientific
explanation, or explanation of any sort.  We might conclude that the
questions that we are asking are relevant, not only to the issue of the
origin of irreversibility, but to most, if not all, problems -- such as
Hume's problem of induction \cite{hume}, of why the future should at all
be expected to resemble the past -- about the nature of science and
the status and justification of scientific theories.  

The point is that in science, as in life, we must learn to cope with
uncertainty. As Boltzmann noted in the above quotation, the mechanical
picture itself is not strictly provable. Nor is any other scientific theory. No
matter how strongly a particular theory seems to be supported by the
evidence at hand, there are always logically possible alternative accounts
of the very same evidence, however far fetched these may be. The account
that we tend to believe -- what we consider to be the inference to the best
explanation -- seems to us simpler, more elegant, more natural, and
certainly less contrived, than the alternatives. Nonetheless, our
expectations for future behavior, based on the best available evidence, can
never be regarded as inevitable consequences of that evidence, but
only -- at best -- as overwhelming likely.

With regard to \lq probablistic' or \lq statistical' explanations,
involving uncertainty about initial conditions, we say that a phenomenon
has been explained if it holds for {\it typical\/} initial conditions, that
is with rare exceptions as defined by a suitable \lq \lq measure'' $\mu$
of  typicality. The phenomenon has been explained if the set $E$ of
exceptional initial conditions satisfies $\mu(E)\ll 1$.

Of course it is essential that the measure of typicality be natural and not
contrived. It should be an object that could, somehow, have been agreed
upon before the phenomenon to be explained was even considered. For
dynamical systems such as we are discussing here, the measure of typicality
should be naturally related to the dynamics, and the most common such
condition is that of stationarity. After all, our notion of typicality
should not change with time. And for classical mechanics, for which
symplectic or canonical structure plays a crucial role in the dynamics, the
most natural measure is the volume measure defined by the symplectic
coordinates, the measure we have been invoking throughout this article.

Here is a small point, but one worth making if we intend to worry a bit
about the justification of explanation via typicality: While typicality is
usually defined -- as it was here -- in terms of a probability measure, the
basic concept is not genuinely probablistic, but rather a less detailed
concept. A measure $\mu$ of typicality need not be countably additive, nor
even finitely additive. Moreover, for any event $E$, if $\mu$ is merely a
measure of typicality, there is no point worrying about, nor any sense to,
the question as to the real meaning of say \lq$\mu(E)=1/2$'. Distinctions
such as between \lq$\mu(E)=1/2$' and \lq$\mu(E)=3/4$' are distinctions
without a difference. 

The only thing that matters for a measure $\mu$ of typicality is
\lq$\mu(E)\ll1$': a measure of typicality plays solely the role of
informing us when a set $E$ of exceptions is sufficiently small that we may
in effect ignore it and regard the phenomenon in question, occurring off
the set $E$, as having been explained. And our future expectation is for
behavior that is typical in this sense. After all, our firm belief in the
impossibility of perpetual motion machines is not based primarily on the
fact that none has ever been achieved so much as on Boltzmann's account of
why such a machine would be a practical impossibility.

What I've just described is the calculus of explanation via appeal to
typicality. The rigorous justification of such a calculus is, as I've
already indicated, another matter entirely, a problem, like so many similar
problems involving crucial uncertainty, of extraordinary difficulty. To
begin to address this problem we would have to investigate how explanation
and typicality are related, which would of course require that we analyze
exactly what is meant by \lq explanation' and by \lq typicality.'  If
typicality is not probability -- the explication of which is itself a
controversial problem of immense difficulty -- it would help to be aware of
this from the outset.

Much has been written about such matters of justification and
nondeductive rationality (for a recent discussion see
\cite{bonjour}). Much more should be written and, no doubt, will be
written. But not here!

%


\begin{thebibliography}{8.}
\addcontentsline{toc}{section}{References}

\bibitem{joel} J.L. Lebowitz: Physics Today \textbf{46}, 32 (1993); \lq
Microscopic Reversibility and Macroscopic Behavior: Physical Explanations
and Mathematical Derivations'. In: \emph{25 Years of Non-Equilibrium
Statistical Mechanics, Proceedings of the Sitges Conference in Barcelona,
Spain, 1994}, Lecture Notes in Physics, ed. by J.J. Brey, J. Marro,
J.M. Rub{\'i}, M. San Miguel (Springer, Heidelberg 1995). For a very recent
comprehensive overview see \lq Microscopic Origin of Irreversible
Macroscopic Behavior', preprint, available on the server at xxx.lanl.gov and
in the Texas Mathematical Physics Archives.

\bibitem{jean} J. Bricmont: \lq Science of Chaos or Chaos in Science?'.
In: \emph{The Flight from Science and Reason}, Annals of the New York
Academy of Sciences \textbf{775}, ed. by P.R. Gross, N. Levitt, M.W. Lewis
(The New York Academy of Sciences, New York 1996) pp. 131--175

\bibitem{Kelvin} W. Thomson: Proceedings of the Royal Society of
Edinburgh \textbf{3}, 325 (1874); reprinted in \cite{brush}

\bibitem{brush} S.G. Brush: \emph{Kinetic Theory} (Pergamon, Oxford 1966)

\bibitem{sc1} E. Schr\"odinger: \emph{What is Life? : The Physical Aspect
of the Living Cell with Mind and Matter \& Autobiographical Sketches}
(Cambridge University Press, Cambridge 1992)

\bibitem{sc2} E. Schr\"odinger: \emph{What is Life? and Other Scientific
Essays} (Doubleday Anchor Books, New York 1965) section~6

\bibitem{b} L. Boltzmann: Annalen der Physik \textbf{57}, 773 (1896);
reprinted and translated as Chapter~8 in \cite{brush} 

\bibitem{gibbs} J.W. Gibbs:  \emph{Elementary Principles in Statistical
Mechanics} (Yale University Press, New Haven 1902) Chapter XII

\bibitem{enm} R. Penrose: \emph{The Emperor's New Mind} (Oxford University
Press, New York, Oxford 1989)

\bibitem{Feynman} R. Feynman: \emph{The Character of Physical Law} (MIT
Press, Cambridge 1967) Section~5

\bibitem{b1} L. Boltzmann: Annalen der Physik \textbf{60}, 392 (1897);
reprinted and translated as Chapter~10 in \cite{brush}

\bibitem{hume} D. Hume: \emph{An Enquiry Concerning Human Understanding}
(Prometheus Books, Amherst, New York 1988)

\bibitem{bonjour} L. Bonjour: \emph{In Defense of Pure Reason: A
Rationalist Account of A Priori Justification},  Cambridge Studies in
Philosophy (Cambridge University Press, Cambridge 1997)

\end{thebibliography}
\end{document}